# Stochastic simulation of binary annihilation reactions within q-analysis formalism

José Bastardo M.[1] and Ney Luiggi A.[1,*]

Grupo de Física de Metales. Dpto. de Física. Escuela de Ciencias. Núcleo de Sucre. Universidad de Oriente. Cumaná. Venezuela.

[*]Corresponding autor

josejbastardom@gmail.com; nluiggi@udo.edu.ve

**ABSTRACT**

Although Gillespie's algorithm is justified under a set of axioms based on the assumption of homogeneity of the system, many chemical systems deviate from this assumption, as is the case for reactions taking place in low-mobility media. Using instead the generalized q formalism, we propose a new stochastic simulation algorithm by redefining the probability with which a µ reaction occurs between $t + \tau$ and $t + \tau + \delta\tau$ as $P(\tau_q,\mu) = a_\mu \exp_q(-a_0 \tau_q)$, taking into account the separation of the natural exponential by the q parameter. Our algorithm has been implemented in the study of binary annihilation reactions, demonstrating a wider amplitude within the range of established physicochemical reactions, being the stochastic Gillespie scheme and the classical deterministic approach particular cases of this new proposal. The effect of the nonextensivity parameter, q, on the reaction rate is analyzed and its relationship with the reaction order, n, and the heterogeneity parameter, h, is determined for two different reactant concentrations in the annihilation reaction. Different behaviors of these parameters are observed for the two types of samples, especially as q moves away from 1, confirming that quasi-second order reactions occur when reactant concentrations are similar and quasi-first order reactions when they are different. Empirical expressions between the classical reaction order and the degree of heterogeneity are proposed. The results obtained allow us to associate the behavior of sub and supra Arrhenius kinetics reported in other different reactions with the degree of nonextensiveness of the reaction.

**RESUMEN**

Aunque el algoritmo de Gillespie se justifica en virtud de un conjunto de axiomas basados en el supuesto de homogeneidad del sistema, muchos sistemas químicos se desvían de este supuesto, como ocurre en las reacciones que tienen lugar en medios de baja movilidad. Utilizando en su lugar el formalismo q generalizado, proponemos un nuevo algoritmo de simulación estocástica redefiniendo la probabilidad con la que se produce una reacción µ entre $t + \tau$ y $t + \tau + \delta\tau$ como $P(\tau_q,\mu) = a_\mu \exp_q(-a_0 \tau_q)$, considerando la separación de la exponencial natural a través del parámetro q. Nuestro algoritmo se ha implementado en el estudio de reacciones de aniquilación binarias, demostrando una mayor amplitud dentro del rango de reacciones fisicoquímicas establecidas, siendo el esquema estocástico de Gillespie y la aproximación determinista clásica casos particulares de esta nueva propuesta. Se analiza el efecto del parámetro de no-extensividad, q, sobre la velocidad de reacción y se determina su relación con el orden de reacción, n, y el parámetro de heterogeneidad, h, para dos concentraciones diferentes de reactantes en la reacción de aniquilación. Se observan comportamientos diferentes de estos parámetros para los dos tipos de muestras, especialmente a medida que q se aleja de 1, lo que confirma que se producen reacciones de cuasi-segundo orden cuando las concentraciones de reactivos son similares y de cuasi-primer orden cuando son diferentes. Se proponen expresiones empíricas entre el orden clásico de reacción y el grado de heterogeneidad. Los resultados obtenidos permiten asociar el comportamiento de las cinéticas sub y supra Arrhenius reportadas en otras reacciones diferentes con el grado de no-extensividad de la reacción.

INTRODUCTION

The study of kinetic processes is a captivating and extremely extensive topic since it covers all fields of science [1-5]. The knowledge of the innumerable ways in which reactants combine to generate products has allowed for the understanding of quotidian phenomena involving simple reactions of atoms and molecules as well as more complex structures in cells, viruses, bacteria, and higher living organisms. The theoretical treatment of the problem comprises two large schemes: The deterministic scheme, where the variables involved in the reaction both set its beginning and, in a classical way, predict its outcome; and the stochastic one, whereby the search of an outcome includes the fluctuations of variables associated with the reaction where the reactant population fluctuates probabilistically, its final state depending on this consideration.

Bimolecular annihilation reactions (A + B → Product), where the reactants do not react or the product fades, have been widely documented as being simple reactions [6] whose solution in both schemes makes it possible to illustrate their limits of coincidence. Considering Tsallis' nonextensive statistics, solutions for this type of reaction change with the degree of nonextensivity, departing from the typical behavior predicted by Boltzmann statistics. This fact was already corroborated for multiple reactions in the works of Aquilanti *et al.* [7] and Luiggi [8]. This paper purports to stochastically study the behavior of binary annihilation reactions within the q-analysis framework.

2. BINARY ANNIHILATION REACTIONS

A chemical reaction of type A + B → $\phi$ is based on the collision of two reactants, A and B, to yield a product that disappears. It is a bimolecular reaction, and the reaction rate equation of species A is given by

$$\frac{d[A]}{dt} = -k[A][B] \qquad (1)$$

where [A] and [B] are the reactant concentrations, and k is the reaction rate constant. For equal initial concentrations of reagents ($[A_0] = [B_0]$), the solution is achieved:

$$[A] = \frac{[A_0]}{1+[A_0]kt} \qquad (2)$$

whereas for [B0] > [A0], the solution is [4],

$$[A] = \frac{[A_0]}{[B_0]e^{-\Delta kt}-[A_0]} \qquad (3)$$

where $\Delta = [B_0] - [A_0]$.

The time required for the concentration of one of the reactants to be reduced by half is called characteristic time τ, given, in the first case, by $(k[A_0])^{-1}$; and in the second, by $\frac{ln|1+\Delta/[B_0]|}{k}$, such that when $[B_0] \gg [A_0] \approx ln|2|(k[B_0])^{-1}$, or the so called pseudo-first-order reaction is achieved. For sufficiently short reaction times (t <<τ) both solutions tend to [A] = [A$_0$] (1- [B$_0$] kt); although for long times, the first reaction tends to decay with $t^{-1}$, and the second tends to decline as $e^{-\Delta\ kt}$. This decay, in general, is characteristic of reactions where there is a large enough number of collisions between A and B per reaction event such that the reaction probability is controlled by the reaction rate constant k [9]. Generally, the behavior of densities at the asymptotic long-time boundary is anomalous, this limit only to be found when the role of fluctuations in the distributions of A and B is considered [10].

The study of the reaction-diffusion systems at low dimensions is abundant, and recently there has been an emphasis on breaking the velocity equations corresponding to the midfield approximation [11]. In fact, these processes follow a power-law format for decay in particle density, as reported by Kopelman [12,13]. This power law poses a tendency to simplify reaction-diffusion equation modeling, as opposed to trends that involve linking the probabilities of multiple particle co-location, where the assumptions imposed [14-16] complicate the accomplishment of a manageable stochastic model. In this work, we will focus our attention on the simulation of a binary annihilation reaction within q-algebra formalism.

The ideas of q-analysis independently developed by Euler, Jackson, Ramanujan, and Watson acquired importance in different fields of science [17], having important applicability in nonextensive statistical physics [18]. Tsallis was the first to introduce it in his generalized entropy function [18]. Recently, Mendes *et al*. [19] considered a q-algebra in the solution of the nonlinear differential equation of n-order kinetics,

$$\frac{dA_n(t)}{dt} = -k_n A_n(t)^n \qquad (4)$$

where parameter n, known as the reaction order, is related to parameter q, indicative of nonextensibility. Equation (4) presents the same functional form as the generic relationship on which Tsallis' nonextensive entropy is based [19],

$$\frac{dp}{dt} = -\lambda_q p^q \qquad (5)$$

When $\lambda_q = \lambda p^{q-1}$, a characteristic first-order reaction is reproduced, p assumes the meaning of an extensive variable in the thermodynamic sense, in accordance with deterministic theory. A detailed analysis of the n or q order kinetic is proposed by Brouers *et al.* [20], who using extended powers and exponential functions defined within a q-algebra, managed to obtain a generalized expression of the so-called 2-parameter fractal kinetics. The proposed solution for equations (4) and (5) is

$$A_q(t) = A(0)exp_q\left(\frac{-t}{\tau_q}\right) = A(0)[1 + (1-q)k_q t]^{\frac{1}{(1-q)}} \qquad (6)$$

with a characteristic time value $\tau_q$ obtained by asymptotic expansion defined by power-law, given by

$$\tau_q = [A(0)^{(1-q)}k_q]^{-1} \qquad (7)$$

$\tau_q$ has a characteristic reaction constant $k_q$ associated with it. These expressions encompass the first-order reaction (q = 1), as well as complex reactions with anomalous kinetics showing behavior in powers of t [21, 22].

McQuarrie [23] presents a stochastic approach to the study of chemical kinetics, translated by Gillespie as a computational algorithm [24-26], in which homogenous chemical kinetics are described as stochastic processes without having to solve the master equation. The procedure bases its effectiveness on the fact that the type of reaction that occurs, as well as the time between consecutive reactions, are stochastic magnitudes. The central hypothesis of this method is that each of the reactions $R_\mu$ ($\mu= 1; 2; 3;..$) would be characterized by a quantity $c_\mu$, where $c_\mu \delta t$ is the average probability for a particular combination of two reactive $R_\mu$ molecules to react in the next interval $\delta t$. If $h_\mu$ represents the total number of combinations of molecules that might react, according to reaction $R_\mu$, the number of reactions per unit of time would be $a_\mu = h_\mu \delta t$. Given a state X at time t, the probability $P(\tau,\mu)d\tau$ for the next reaction $R_\mu$ to occur in an interval [t + τ and t + τ + δτ] is given by:

$$P(\tau,\mu)d\tau = P_0(\tau)a_\mu d\tau \qquad (8)$$

where $a_\mu d\tau$ is the probability that a reaction will occur in the time interval [t + τ and t + τ + δτ]; and $P_0(\tau)$, the probability that no reaction will occur in this same interval. If the nonreactive collisions between the molecules are to be much more frequent than the reactants, and allow for a uniform distribution of the molecules throughout the volume, then the particle velocities will be randomly distributed according to the Maxwell-Boltzmann distribution, such that it is possible to use:

$$P_0(\tau) = exp(-a_0\tau) \qquad (9)$$

which permits to write Equation (8) as

$$P(\tau,\mu) = a_\mu exp(-a_0\tau) \qquad (10)$$

Expression (10) defines the probability that reaction $R_\mu$ will occur in the system within time interval τ, the objective behind this computational method being to obtain τ by means of the Monte Carlo simulation. For that, the cumulative integral of Equation (10) must first be calculated, the time remaining to be set by

$$\tau = \frac{-ln(r_1)}{a_0} \qquad (11)$$

where $a_0$ does not explicitly depend on time, and $r_1$ is a random number evenly distributed in interval (0, 1). Equation (11) is used to obtain the time it takes for the next reaction to occur in a homogeneous system [25,27]. The advantage of the Gillespie computational method is its easy and practical implementation. Besides, it permits to reproduce the behavior of the stochastic system with just a few particles, although it cannot handle a heterogeneous environment. We propose to extend this formalism to anomalous systems, supported by q-algebra.

3. APPLICATION OF q-ALGEBRA TO THE STOCHASTIC FORMULATION

It is proposed that in an anomalous system, the probability that the following reaction may occur in (t + τ, t + τ + δτ) is given by:

$$P(\tau_q,\mu) = a_\mu exp_q(-a_0\tau_q) \qquad (12)$$

As before, a new time τ is sought using the Monte Carlo method, where the cumulative integral must first be calculated using the probability proposed in Relation (12), such that

$$r_1 = \int_0^{\tau_q} P(\tau,\mu)\, dt = \int_0^{\tau_q} a_\mu exp_q(-a_0\tau)\, dt \qquad (13)$$

and using the properties of q-algebra [15], we get

$$r_1 = \frac{a_\mu}{(2-q)a_0}\left[1 - exp_q(-a_0\tau_q)^{2-q}\right] \qquad (14)$$

Considering $P(\tau,\mu) = P(\mu)P(\tau)$, Equation (14) is written as

$$r_1 = \frac{1}{(2-q)}\left[1 - exp_q(-a_0\tau_q)^{2-q}\right] \qquad (15)$$

Different generalizations of the classical exponential can be found in the literature [28]. In particular, a significant amount of work has been developed [28-31] about the q-exponential function and its distribution, whose application has led to a novel interpretation of statistical physics, specifically in the theory of reactions that depart from traditional Arrhenius behavior.

In this work, we use the following expressions: In the neighborhood of q = 1,

$$exp_q(x) = \begin{cases} (1 + (1-q)x)^{\frac{1}{1-q}} & para \quad 1 + (1-q)x > 0 \\ 0 & para \quad 1 + (1-q)x \leq 0 \end{cases} \qquad (16)$$

Or

$$exp_q(x) = [1 + (1-q)x]_+^{\frac{1}{(1-q)}} \qquad (17)$$

where $[a]_+$ means max $\{0, a\}$ and the parameter q is a number in the range $0 < q < 2$.

A new expression for $\tau$ is obtained

$$\tau_q = \frac{1-r_2}{(1-q)a_0} \quad \text{where } r_2 = \begin{cases} (1-qr_1)^{\frac{1-q}{q}} & for \quad 0 < q < 1 \\ 1 - (2-q)r_1^{\frac{1-q}{2-q}} & for \quad 1 < q < 2 \end{cases} \qquad (18)$$

Using heuristic reasoning, we define the propensity, demanding the classical propensity outcome when q = 1.

It is known that the propensity for a bimolecular reaction is $a_0 = k[A][B]$ for the Gillespie case, while for the case of anomalous diffusion (super-diffusion) the following propensity is proposed:

$$a_0 = k_q[A]^{2-q}[B]^{2-q} \qquad (19)$$

where $k_q = k[A_0]^{q-1}[B_0]^{q-1}$ for $0 < q < 1$; however, for the other case (sub-diffusion), it must be fulfilled that

$$a_0 = k_q[A]^q[B]^q \qquad (20)$$

where $k_q = k[A_0]^{1-q}[B_0]^{1-q}$, such that when $q \to 1$, the classical propensity is recovered for both cases. The rest of the rules of the evolution of the Gillespie algorithm remain without any change.

## 4. RESULTS AND DISCUSSION

Binary annihilation kinetics will be evaluated for each of the schemes developed only for two concentrations of particular reactants ($A_0 = B_0$ and $A_0 < B_0$), highlighting in each case the effect on the characteristic time and the parameter of nonextensivity q.

Prior to choosing the number of particles with which the present study would be carried out, we checked our algorithm with 500, 1,000, 5,000 and 10,000 particles, obtaining similar results on average, with a greater fluctuation as the sample was smaller; and obviously, increasing the calculation time with the growth of the sample, thus confirming that the Gillespie algorithm performs well for small samples. It is for these reasons that a sample size of 1,000 particles has been selected for this study. The behavior of our formulation around $q = 1$ and the convergence of the different models to the classic model by far justify the applicability of our approach.

### 4.1 For $[A_0] = [B_0]$

Our first study relates to samples with equal numbers of reactant particles, where the classical model is represented by Eq. (2). Figure 1 shows the decay of the number of particles A as a function of time on both a linear scale (Fig. 1.a) and a log-log scale (Fig. 1.b) for a random run at different q values.

This figure shows an Arrhenius-type behavior associated with classical or mean-field kinetics, which agrees with Gillespe's kinetics throughout the calculation for values of q equal to or very close to 1, although very small deviations are detected at the beginning of the reaction. In addition, the simulation provides details on the effect of nonextensivity on the kinetics of the reactants, with these kinetics falling below the classical kinetics as we move away from q=1, for times well below the time required for the reaction to consume half of the reactants ($t \ll t_{1/2}$). This means that at the beginning of the simulated reaction, its behavior is of the sub-Arrhenius type. After this time ($t > t_{1/2}$), a transition to values above classical kinetics is observed, with a larger deviation for values of q further away, above unity. These results suggest an anomalous asymmetric (sub-super Arrhenius) kinetic behavior of the q-dependent reaction with respect to classical kinetics.

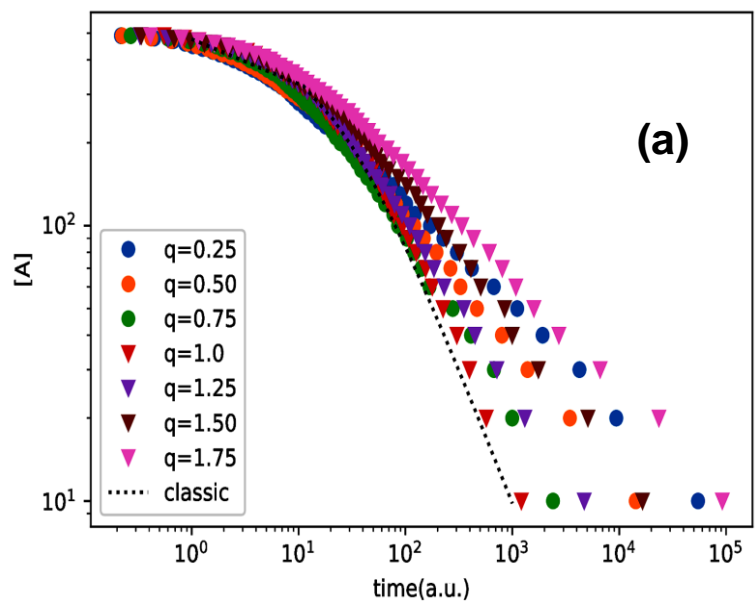

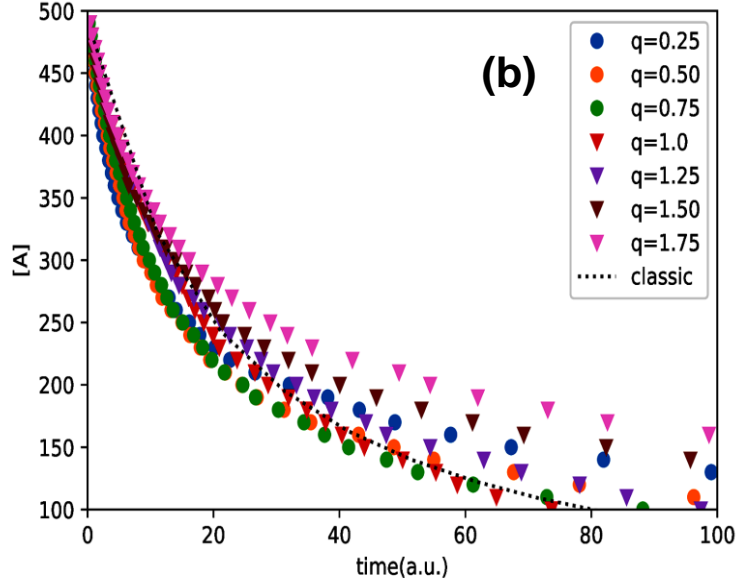

Figure 1: Concentration of particles [A] *vs* (time), with k = 1.10$^{-4}$ for $A_0 = B_0 = 500$. The dotted black curve is obtained from Equation (2) in the midfield approximation, under the same conditions. (a) On a linear scale. (b) On a log-log scale.

The characteristic behavior of the anomalous systems for long times follows the relation $A \sim A_0 t^{-(1-h)}$, verifying that when q → 1, a behavior close to the mean field is reproduced; that is, h → 0. h increases when q moves away from 1 (see Table 1).

4.1.1. Reaction rate

Considering that the reaction rate for a random run presents large fluctuations inherent to the method, but always fluctuating around the mean values, then, a good approximation to the global behavior of this parameter is achieved by using the definition given by Gillespie:

$$\frac{dA(t)}{dt} \cong \frac{1}{\tau} \approx a_0 A(t) B(t) \quad (21)$$

However, the fractal-like equation for the condition $[A_0] = [B_0]$ remains as

$$\frac{dA(t)}{dt} = k_q(t)[A(t)]^2 \quad (22)$$

which is equivalent to Equation (4) and tends to the mean field equation

$$\frac{dA_n(t)}{dt} = k_n A_n(t)^n \quad (23)$$

where $k_n$ is a constant and the n value is evaluated from the slope of the graph ln (dA / dt) vs ln (A). The heterogeneity degree of the reaction will be characterized by the n order.

By plotting the variation of the logarithm of the reaction rate, ln (dA / dt) versus ln (A), for different values of q, as shown in Figure 2, these values are determined. A straight line is obtained for each q, indicating that the reaction rate follows a power law of A.

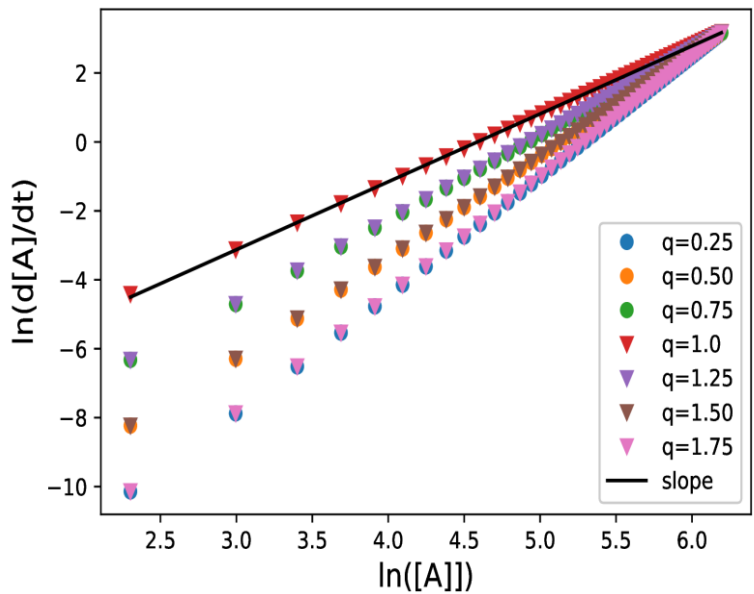

Figure 2: ln (dA / dt) vs ln (A) with k = $1.10^{-4}$ for $A_0$ = $B_0$ = 500. The black curve represents the reaction speed for q = 1, with n = 1.97.

Figure 3 plots the behavior of n and of the constant $k_n$ with q, n showing a linear decrease with q for q < 1 and a linear growth for q > 1, while the logarithm of the reaction constant shows a linear but inverse behavior of n. In both cases, proportionality between n and ln $k_n$ with q is established. These graphs clearly show that the order of kinetics and the logarithm of the reaction constant depend linearly on the degree of nonextensiveness when the spontaneous annihilation reaction occurs between reactants in equal proportions.

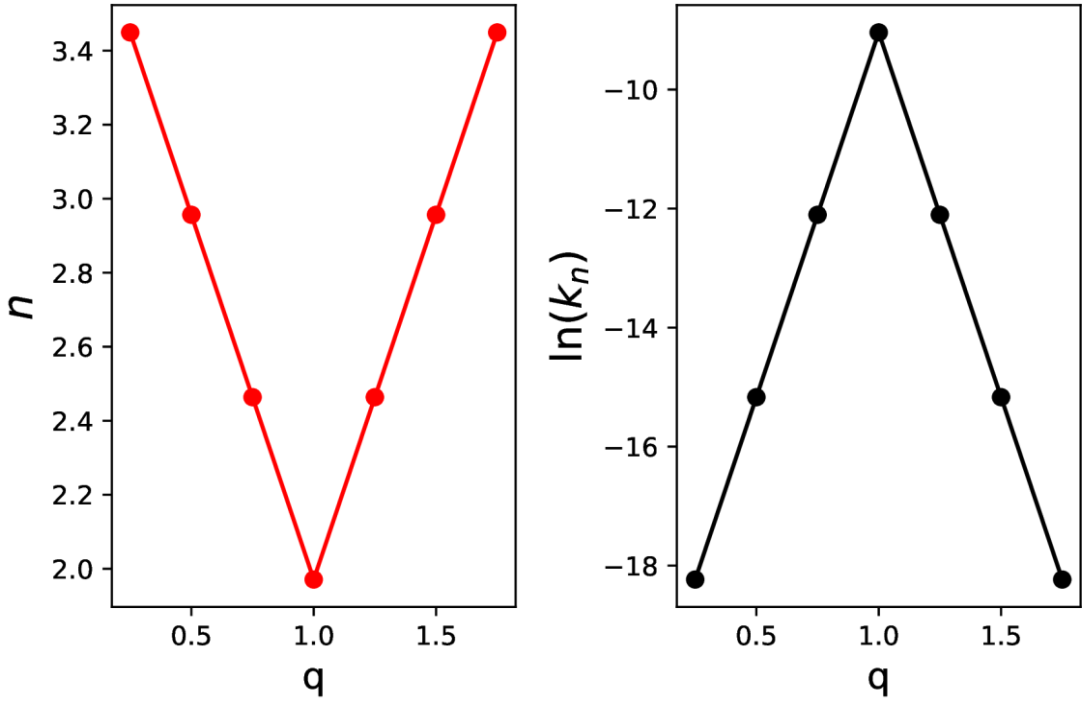

Figure 3: Variation of $k_n$ and n parameters with q

4.1.2 Fractal Kinetics in Q-Formalism.

The chemical kinetics of heterogeneous disordered systems is extremely complicated due to the energetic and geometrical complexity that underlies it, but heterogeneous reactions are present in most reactions not only at the atomic and molecular level, but also at the level of photophysical annihilation of excitations, where the usual volumetric diffusion length of homogeneous reactions must be replaced by the half-random paths of these excitations, which opens a suitable referential framework for the study of low-dimensional kinetics. Many events of this type are mentioned in the literature: catalysis in porous materials, biochemical reactions in membranes, electron-hole recombination on the surfaces of amorphous materials, trapping and annihilation of excitations, biomedical reactions in inhomogeneous media, etc., where the dominant mechanisms in the kinetics of these events show a fractal behavior, controlled by annihilation reactions (A+A → Product), and whose reaction rate in a system of similar reactants is represented by Eq. 22.

Kopelman [12] in his study "Rate Processes on Fractals: Theory, Simulations, and Experiments" shows that for diffusion-limited reactions the effective reaction coefficient k(t) follows a power law for long times,

$$k(t) \sim k_0 t^{-h} \qquad (24)$$

Where h, the heterogeneity exponent, measures the degree of heterogeneity of the reaction and is related to the effective spectral or fractal dimensionality by h=1- ds/2.

The simulation of the reaction gives us access to the reaction rate through Eq. 22, while using Eq. 24 we graphically determine the relationship between the parameters q and h, as shown in Figure 4(a), where the evolution of k(t) is plotted as a function of time. The heterogeneity parameter (h) for different values of the nonextensivity parameter q is obtained from the slope of this plot. It is observed that as q changes, so do the slopes at long times, although the linearity suggested in reference [12] for short times is broken, indicating a possible instability at the beginning of the reaction. Similar behavior has been observed in the early stages of homogeneous solid nucleation reactions..

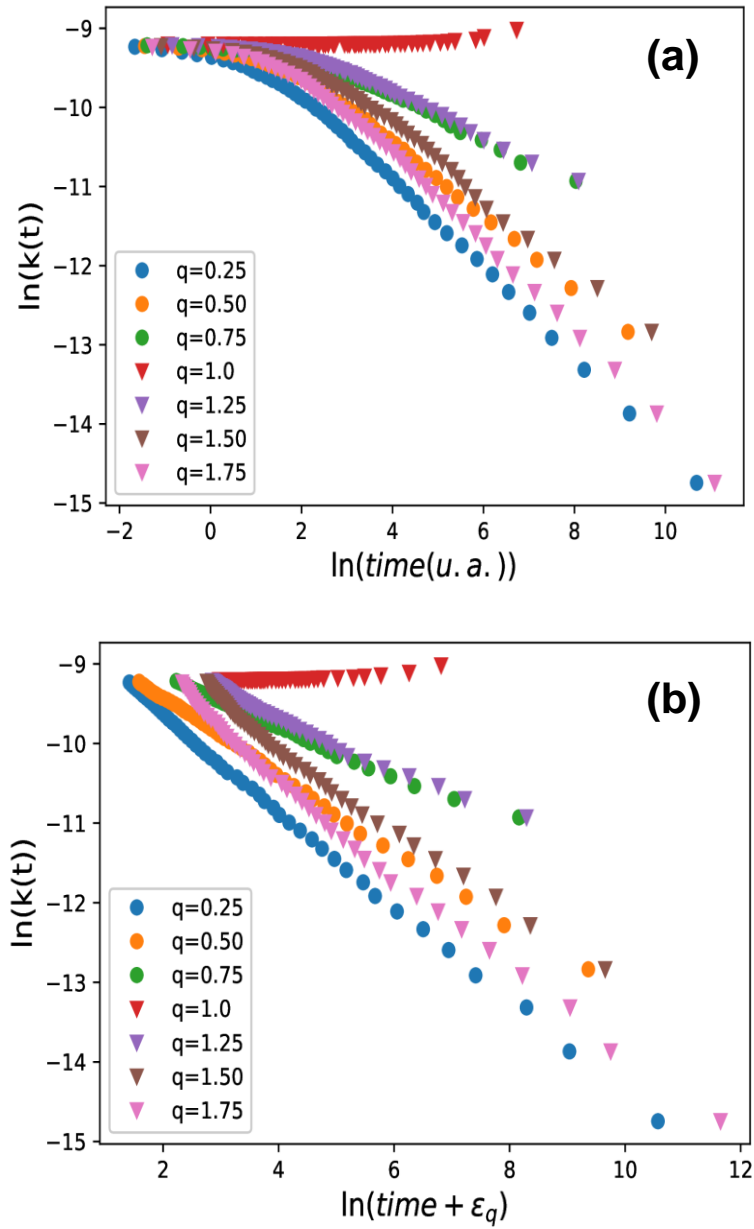

Figure 4. (a): ln k(t) vs ln time with. (b) ln k(t) vs ln (time+ξq). With A0 = B0 = 500, $k_0$ = $1.10^{-4}$. ξq = {4.0, 4.7, 9.0, 19.0, 17.0, 15.4, 10.0} and q = {0.25, 0.50, 0.75, 1.0, 1.25, 1.50, 1.75}, respectively.

Schnell and Turner [21] suggest the use of a more general equation for *k(t)*, which does not diverge for very short times,

$$k(t) = k_0(t + \xi)^{-h} \qquad (25)$$

where *k₀* and ξ, in principle, are considered adjustment parameters. The scale parameter ξ corresponds to the time where the anomalous effects begin to be seen. This parameter has no meaning for the classical case. To facilitate mathematical manipulation, the time scale ξq in Equation (25) is powered by the exponent −h, which is evaluated from the log-log graph of *k(t)* vs (t + ξq); the values of ξq are selected in such a way that the graph in question is linearized. Figure 4.b shows the linearization of the curves with the respective ξ parameters. The value of h is obtained from the slope of each line (see Table 1). The behavior of the parameters h as a function of the nonextensivity parameter q is shown in Figure 5, where a departure from linearity is observed as we move away from the mean field condition.

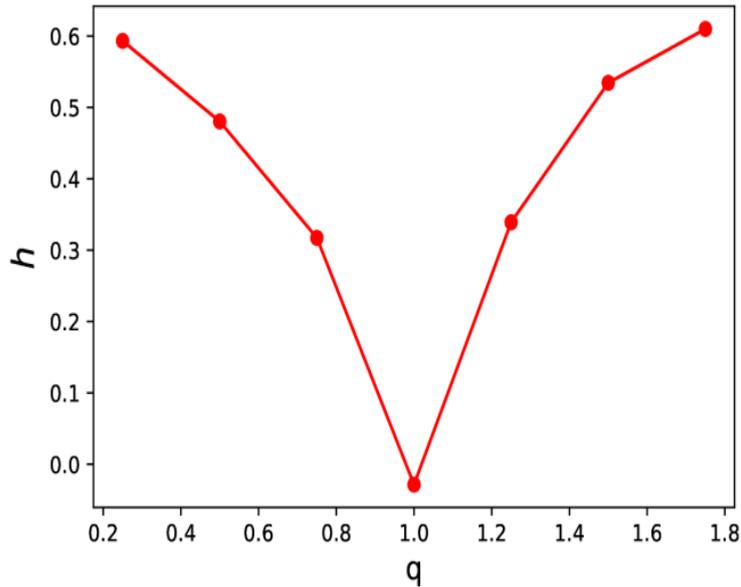

Figure 5. h vs q obtained from the linearization of Figure 4.b

Our results also suggest an empirical relationship between the reaction order (n) and the heterogeneity exponent (h), given by

$$n = \frac{2-h}{1-h} \qquad (26)$$

At the end of this section, the parameters derived from the simulation of the annihilation reaction for [A0] = [B0] are summarized in Table 1.

Table 1. Parameters obtained for $[A_0] = [B_0]$. Exponent n, h, time scales $\xi_q$, $t_{1/2}$ and $t_{1/4}$ for different q's

| q | n | h | $\xi_q$ | $t_{1/2}$ | $t_{3/4}$ |
|---|---|---|---|---|---|
| 0.25 | 3.44 | 0.60 | 4.0 | 15.67 | 54.8 |
| 0.50 | 2.9 | 0.47 | 4.5 | 17.77 | 42.7 |
| 0.75 | 2.46 | 0.30 | 9.0 | 17.05 | 37.2 |
| 1.0 | 1.97 | 0.02 | - | 21.45 | 41.3 |
| 1.25 | 2.46 | 0.33 | 17.0 | 22.80 | 50.9 |
| 1.50 | 2.95 | 0.53 | 15.4 | 26.45 | 73.4 |
| 1.75 | 3.44 | 0.60 | 10.0 | 30.29 | 98.0 |

Table 1 shows the characteristic parameters of the simulation for $[A0] = [B0]$. The parameters n and h show a symmetric behavior with respect to the values derived from the mean-field theory, corresponding to q=1, n=1.97 and h=0.02. The time scale tends to increase asymptotically as we approach q=1, this increase being greater for values of nonextensivity greater than 1. The characteristic time as a function of the parameter q is also shown when 1/2 and 3/4 of the reaction has elapsed. These times reflect different characteristics of the kinetics as the reaction progresses, with the values of t1/2 and t3/4 showing slower and monotonically decreasing kinetics for q > 1 with respect to the kinetics predicted by the mean-field approximation. However, for q < 1 some randomness is observed, suggesting the presence of the short-time superdiffusion phenomenon characteristic of some anomalous systems, which causes some half-lives to be shorter than the classical half-life. At later stages of the reaction and q < 1, the kinetics become faster around q = 0.75, which assumes a definite functional form due to the tendency of the system to stability, then the reaction tends to slow down as the value of nonextensivity increases. This table also confirms that a quasi-second-order reaction, typical of chemical adsorption reactions, takes place as we move away from the mean-field approximation (away from q=1).

The results shown above, for this condition, represent an important advance in projecting, both above and below the classical deterministic results, a way of explaining the sub-Arrhenius and supra-Arrhenius reaction kinetics reported extensively in the literature.

### 4.2 For $[A_0] < [B_0]$

In this section, the modified Gillespe model is used to simulate the evolution of the annihilation reaction, for a system of 1000 particles, 200 of $A_0$ and 800 of $B_0$, considering different values of the nonextensivity parameter q.

Figure 6 shows the progress of the reaction through the evolution of the concentration of minority reactant A as a function of time at different q's. In 6.a the entire reaction is shown on a log-log scale, while in 6.b, only the beginning of the reaction is shown on a linear scale. A tendency to reproduce the behavior of the classical kinetics for very short reaction times (t < 2 a.u.) is observed, independently of the value of q. In any case, the Gillespe simulation tends to reproduce a sub-Arrhenius behavior in the first instants of the reaction, remaining so except for q=1, whose simulated kinetics is very slightly above the classical kinetics and then moves below it for longer times. In the final stages of the reaction, a super-Arrhenius behavior is observed for all values of q other than one. This changing behavior around the classical kinetics, determined by the mean-field approximation, predicts an anomalous behavior characteristic of systems with low mobility at long reaction times.

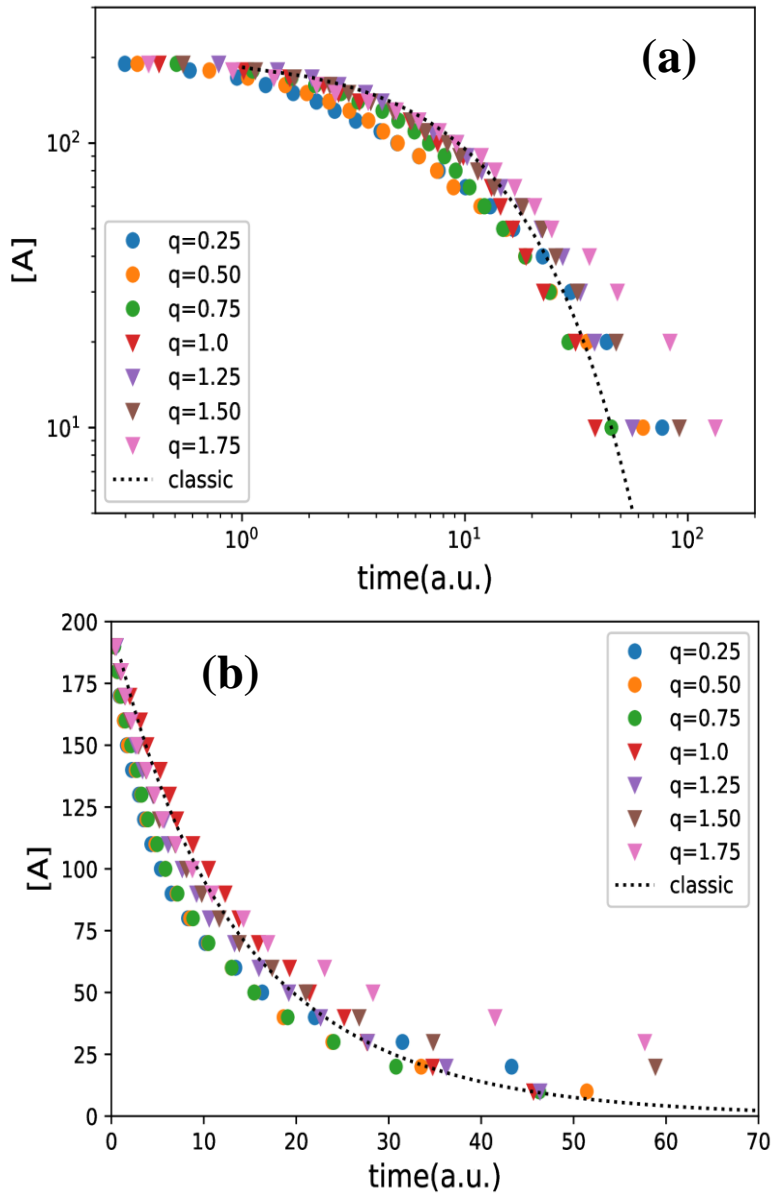

Figure 6. Variation of concentration [A] versus time, for $A_0 = 200$ and $B_0 = 800$. $k_0 = 1.10^{-4}$. The dashed black curve is obtained from equation (4) in the mean-field approach. 6.a. Whole reaction in log-log scale. 6.b. Start of reaction in linear scale.

The deviation of the simulated kinetics from the behavior determined by the mean-field approach affects the characteristic times, whose variation depends on the nonextensivity parameter q.

Table 2 shows the characteristic time when ½ and ¾ of the reaction has occurred for different values of the nonextensivity parameter. Contrary to the case of A0=B0, the characteristic half-life t1/2 and t3/4 show a random variation around q=1, making it difficult to detect a pattern.

4.2.1. Reaction rate

Figure 7 shows the reaction rate versus the number of particles of the minority species (A) for different q's. Using Equation (23), we can find the order of the kinetics represented by exponent n by evaluating the slope of the lines shown for that figure.

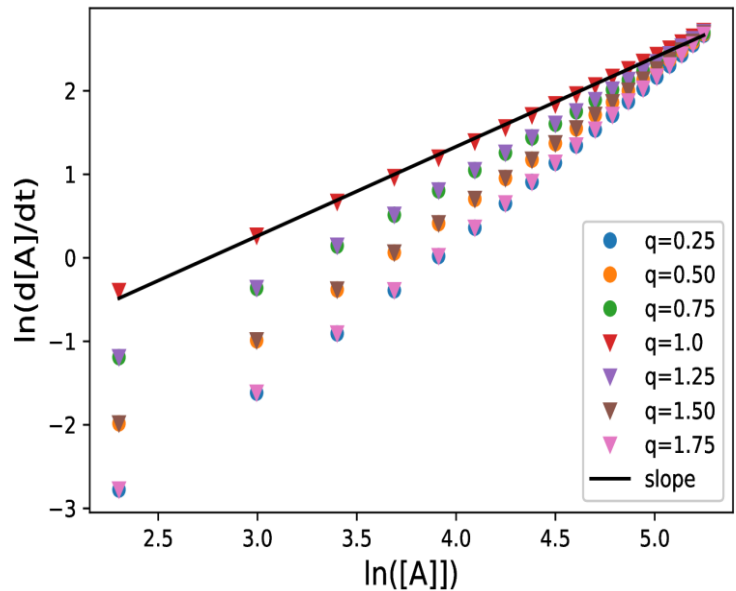

Figure 7: ln (dA / dt) vs ln (A) with k = $1.10^{-4}$ for $A_0$ = 200 and $B_0$ = 800. The black curve represents the reaction speed for q = 1, with n = 1.06.

The variation of the reaction order n is shown in Figure 8; its behavior is qualitatively similar to that of the previous case, although the magnitudes are different. This result allows us to conclude that, independent of the concentration of the reactants, there is a linear decreasing dependence of the reaction order n on the nonextensivity parameter q when q<1 and an increasing dependence for q>1.

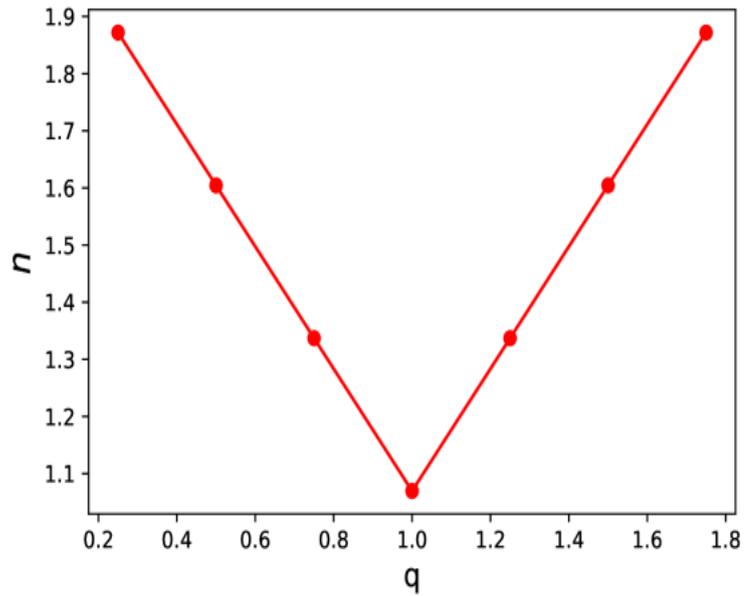

Figure 8. Behavior of the reaction order n vs q using k = $1.10^{-4}$, $A_0$ = 200, and $B_0$ = 800.

As for Kolpman's heterogeneity exponent h [12] and Schnell and Turner's scaling parameter ξq [21], their evaluation is performed in the same way as in the previous case. Figure 9.a shows the evolution of the velocity coefficient *k (t)* vs. time. The figure shows the effect of nonextensivity on the reaction heterogeneity exponent, confirming that the slope h changes significantly with q for long reaction times.

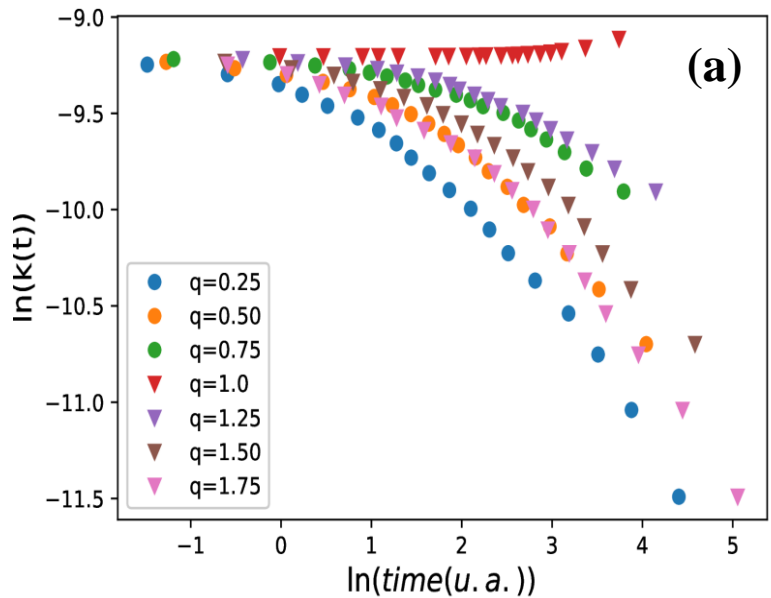

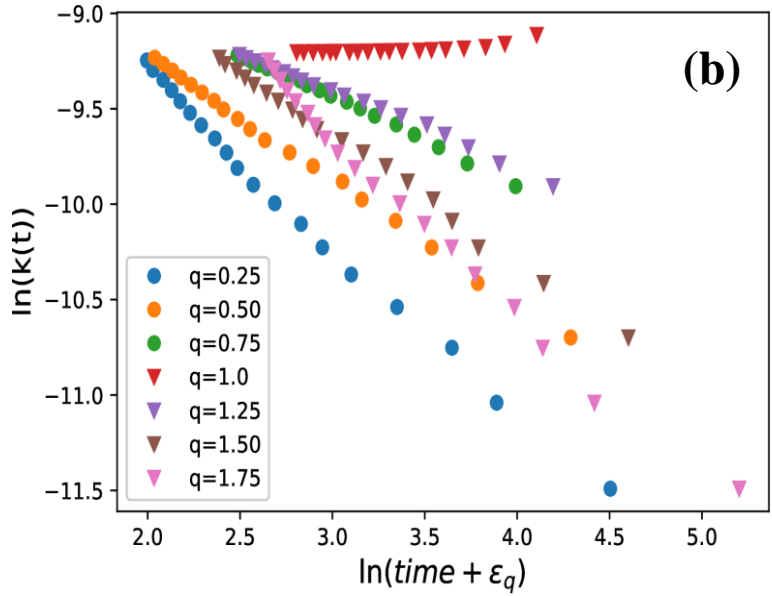

Figure 9: (a) ln k(t) vs ln time (b) ln k(t) vs ln (time+ξq). With $k_0 = 1.10^{-4}$, $A_0 = 200$, and $B_0 = 800$ Figure 10. Ln k (t) vs ln (time + ξq) where ξq = {7.0, 7.3, 11.6, 15.5, 11.7, 10.5, 13.7}, for the respective q values..

As we did in the $A_0 = B_0$ case, exponent h will be found using the evolution of *k (t)* vs (t + ξq) on a log-log scale. In it, as in the previous case, an anomalous decay of the number of particles is observed, where the reaction coefficient falls as (t + ξ) – h. The results of the linearization are shown in Figure 9.b.

On the other hand, Figure 10 shows the dependence of h on q, showing some randomness for q values far from q = 1.

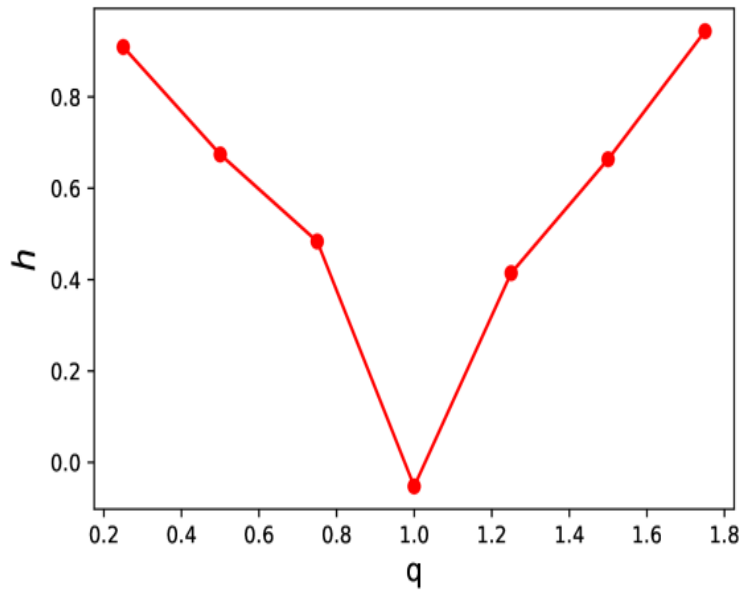

Figure 10: h vs q with $k = 1.10^{-4}$ for $A_0 = 200$ and $B_0 = 800$.

The values of the parameters deduced from the simulation under this condition are shown in Table 2.

Table 2. Parameters obtained for $[A_0] = 200$ and $[B_0] = 800$.
Exponents n, h, time scales $\xi q$, $t_{1/2}$ and $t_{3/4}$ for different q's

| q | n | h | $\xi q$ | $t_{1/2}$ | $t_{3/4}$ |
|---|---|---|---|---|---|
| 0.25 | 1.87 | 0.89 | 7.0 | 5.33 | 16.52 |
| 0.50 | 1.60 | 0.65 | 7.3 | 5.83 | 15.34 |
| 0.75 | 1.33 | 0.45 | 11.6 | 5.85 | 15.27 |
| 1.00 | 1.06 | 0.04 | - | 10.51 | 21.72 |
| 1.25 | 1.33 | 0.39 | 11.7 | 7.66 | 19.23 |
| 1.50 | 1.60 | 0.66 | 10.5 | 8.12 | 21.02 |
| 1.75 | 1.87 | 0.90 | 13.7 | 8.76 | 28.50 |

Table 2 shows that for this ratio between reactant concentrations (B0 >> A0) the reaction behaves as a pseudo first order reaction of B with respect to A, where the evolution of the concentration gradient is clearly not constant. This behavior has been observed in processes involving physical adsorption reactions. Despite the difficulty of establishing a relationship between the reaction order n and the heterogeneity exponent h for pseudo first order reactions, we propose the following empirical relationship from the data obtained from the simulated kinetics:

$$n = \frac{2-h}{2(1-h)} \qquad (26)$$

which is well-behaved around values of q = 1, *i.e.*, the classical case.

5. CONCLUSIONS

We have studied the behavior of the characteristic parameters associated with a binary annihilation reaction within the framework of the Gillespie theory and Tsallis's nonextensive statistical theory, concluding the following:

- The greater breadth of the new algorithm is evidenced, with the Gillespie stochastic scheme and the classical deterministic approach being particular cases of this new proposal.

-The supra and sub Arrhenius reaction kinetics are completely covered by this scheme.

- The same Gillespie structure to find the next reaction that will occur is maintained, which allows us to infer that it is easily applicable to heterogeneous systems with multiple stages.

- The effect of the nonextensivity parameter, q, on the reaction rate is analyzed and its relationship with the reaction order, n, and the heterogeneity parameter, h, is determined for two different reactant concentrations in the annihilation reaction. Different behaviors of these parameters are observed for the two types of samples, especially as q moves away from 1, confirming that quasi-second order reactions occur when reactant concentrations are similar and quasi-first order reactions when they are different.

- The transition time $\xi q$, where the effects of heterogeneity begin to be seen, is dependent on parameter q. This scale generally decreases as q decreases, a more detailed study of this parameter being necessary to know its true nature.

- Empirical relationships are established between the order of the reaction and the coefficient of heterogeneity.

ACKNOWLEDGMENTS

The authors thank Carlos Mota for his support and translation of the manuscript.